\magnification=1100

\hsize 17truecm
\vsize 23truecm

\font\twelvec=msbm10 at 12pt
\font\sevenc=msbm10 at 9pt
\font\fivec=msbm10 at 7pt

\newfam\co
\textfont\co=\twelvec
\scriptfont\co=\sevenc
\scriptscriptfont\co=\fivec

\def\deg{\mathop{\rm deg}\nolimits}
\def\det{\mathop{\rm det}\nolimits}
\def\exp{\mathop{\rm exp}\nolimits}

\def\Id{\mathop{\rm Id}\nolimits}

\def\ker{\mathop{\rm Ker}\nolimits}
\def\lim{\mathop{\rm lim}\nolimits}

\def\Tra{\mathop{\rm Tr}\nolimits}
\def\tra{\mathop{\rm tr}\nolimits}

\def\ssl{\mathop{\rm sl}\nolimits}
\def\Ssl{\mathop{\rm SL}\nolimits}

\def\Sum{\displaystyle\sum}

\baselineskip 15pt

\centerline{\bf QUANTUM VORTICITY AT THERMAL EQUILIBRIUM FOR SPINS SYSTEMS}

\centerline{\bf WITH CONTINUOUS SYMMETRY}
\medskip
\centerline{D.MINENKOV ${}^{(1)}$ and M.ROULEUX ${}^{(2)}$}
\medskip
\centerline {${}^{(1)}$ Institute for Problems in Mechanics of Russian Academy of Sciences}

\centerline {Prosp. Vernadskogo 101-1, Moscow, 119526, Russia, minenkov.ds@gmail.com}
\medskip
\centerline{${}^{(2)}$ 
Aix Marseille Universit\'e, CNRS, Centre de Physique Th\'eorique, UMR 7332, 13288 Marseille, France} 

\centerline {\& Universit\'e de Toulon, 
CNRS, CPT, UMR 7332, 83957 La Garde, France, rouleux@univ-tln.fr}               
\bigskip
\noindent {\bf Abstract}: We propose a definition of vorticity at inverse temperature $\beta$ for Gibbs states in quantum XY spin systems
on the lattice by testing $\exp[-\beta H]$ on
a complete set of observables (``one-point functions''). We show in particular that it is independent of the choice of a
particular basis. Imposing a compression of Pauli matrices
at the boudary, which stands for the classical environment, we make some numerical simulations on finite lattices,
and exhibit usual vortex patterns.
\bigskip
\noindent {\bf 0. Introduction}.
\medskip
Consider the quantum XY or Heisenberg spin model for $S=1/2$ on the 2-D lattice ${\bf Z}^2$, with nearest neighbor interactions.
Marmin-Wagner, and Hohenberg theorems tell that Gibbs states, for all inverse temperature $\beta$, are invariant
under simultaneous rotation of spins (absence of continuous symmetry breaking in two dimensions).
In the classical case, we know a bit more~: although there is a unique Gibbs state, with rotational symmetry,
which rules out the existence of first order transitions, a particular form for phase transition exists,
characterized by a change of behavior in the correlation functions. For the XY system, it has been
described by Berezinskii, and Kosterlitz-Thouless in term of topological excitations, called vortices [FrSp].
For the Heisenberg model, we observe higher order topological defects, called instantons  [BePo].
See also [El-BRo1,2] for the classical Kac's model.

This paper is a first attempt to answer the natural question~: {\it How to define  vortices in the quantum case ?}
Consider first a system in finite volume $\Lambda\subset{\bf Z}^2$.
The Hamiltonians are of the form $H_\Lambda(\Phi)=-\Sum_{X\subset\Lambda}\Phi(X)$, where
$\Phi$ is an ``interaction'' between sites in $\Lambda$.
For nearest neighbor interaction, the contributing $X$ are pairs $\langle i,j\rangle$, and the Hamiltonian of the form
$$H_\Lambda=-{1\over2}\Sum_{\langle i,j\rangle\subset\Lambda}(\sigma_i^x\otimes\sigma_i^x+\sigma_i^y\otimes\sigma_i^y)\leqno(0.1)$$
possibly adding the linear term $\Sum_{i\in\Lambda}\sigma_i^z$. We can also glue to $\Lambda$ a finite boundary $\partial\Lambda$ and modify 
$H_\Lambda(\Phi)$ to $H_{\Lambda\cup\partial\Lambda}(\Phi)$ accordingly, accounting for an approximate ``external field''. 
In finite volume $\Lambda\cup\partial\Lambda$, the only (normalized) Gibbs state is given by
$$A\mapsto\omega_\beta(A)={\tra (e^{-\beta H_{\Lambda\cup\partial\Lambda}} A)\over \tra (e^{-\beta H_{\Lambda\cup\partial\Lambda}}) }\leqno(0.2)$$
and called  the ``canonical Gibbs state''.  We shall actually define vorticity at inverse temperature $\beta$
by decomposing the linear form $\omega_\beta$ on a canonical (orthonormal) basis of observables.

Indeed, to favour the existence of
vortices in finite volume, we have completed $\Lambda$ by a finite volume environment $\partial\Lambda\subset{\bf Z}^2$
where Pauli matrices $\sigma_j$ are ``compressed''
in given directions $(\theta_j)_{j\in\partial\Lambda}$.
Thus, measurements of the observable ``direction of spin''
are deterministic on $\partial\Lambda$, which accounts for the  ''classical'' aspect of the environment, but the
sign of spin still remains a free variable, allowing for a certain ``chessboard symmetry'' of the Hamiltonian,
and comparison between ferro- and antiferromagnetic behaviors.
More precisely, for
$j\in\partial\Lambda$ we replace $\sigma_j$ by
$\sigma_j(\theta_j)=\Pi_{\theta_j}\sigma_j\Pi_{\theta_j}$ where $\Pi_{\theta_j}$ is the orthogonal projection in the
direction ${\cos\theta_j\choose \sin\theta_j}$.

One of the main requirements for consistency of our definition is to check co-variance of the vorticity matrix
with respect to the choice of orthonormal basis.

This paper is organized as follows:

In Part 1, we define vorticity matrices as the decomposition of the Gibbs state in a certain orthonormal basis of 1-point functions;
we call {\it quantum vortices} the points of the lattice where the vorticity matrix vanishes.

In Part 2, we study holonomy properties on the Lie group $\Ssl(2;{\bf R})$, and show how to define
(in the ideal continuous limit) the class of holonomy of vorticity matrices through the ``non-commutative degree''.

In Part 3, we carry some numerical simulations on constrained quantum anisotropic
XY systems, and provide pictures of vortices at thermal equilibrium.

In the Appendix we show that vorticity matrices are defined intrinsically, i.e.
don't depend, up to unitary equivalence, on the choice of a specific orthonormal basis
within a simple class of observables.

To some extend, our approach can be generalized to Heisenberg model, replacing vortices by ``instantons'', or ``skyrmyons''
as in Belavin-Polyakov theory.
But here the non-commutative calculus makes difficult to define properly the degree. This will be hopefully investigated in some later work. 
\medskip
\noindent{\bf Acknowledgments}: We thank A.Messager and S.Shlosman for interesting discussions. This work was initiated in
Hicham El-Bouanani's PhD Thesis [El-Bo] at Toulon University, and 
results in Appendix were obtained with Renaud Ruamps in his unpublished Master's Thesis [Ru] at Aix-Marseille University, 
both under supervision of the second author. 


\bigskip
\noindent {\bf 1. Vorticity matrices}
\medskip
Gibbs state (0.1) for spin 1/2 systems, as a linear form on the ${\bf C}^*$-algebra of observables
$${\cal O}=\otimes_{j\in\Lambda\cup\partial\Lambda}o_j, \quad  o_j={\cal M}_{2\times2}({\bf C})$$
(``quasi-local observables'' if we were to consider the
thermodynamical limit,~) can be decomposed in a canonical basis. The simplest way is to restrict
to ``one-point functions'', i.e. the set $\widetilde{\cal O}\subset{\cal O}$
of $2N\times2N$, block-diagonal $2\times2$ matrices, supported on individual sites of
$\Lambda\cup\partial\Lambda$,  $N=|\Lambda\cup\partial\Lambda|$.
More specifically, let again
$\widetilde{\cal O}_{\bf R}\subset\widetilde{\cal O}$ be a real sub-algebra
$\widetilde{\cal O}$, of real dimension $4N$.
\medskip
\noindent {\it Example 1}:
$\widetilde{\cal O}_{\bf R}$ is the ``canonical'' algebra, generated by real matrices $(D^i)_{i\in\Lambda\cup\partial\Lambda}$,
whose all non-diagonal $2\times2$ blocks vanish, and all diagonal $2\times2$ blocks vanish, except
this supported on site $i$ that takes values in
$\{\delta_1,\delta_2,\delta_3,\delta_4\}$, where
$$\delta_1=\pmatrix{1\kern 1pt &0\kern 1pt \cr
0\kern 1pt & 0\kern 1pt }, \quad \delta_2=\pmatrix{0\kern 1pt &1\kern 1pt \cr
0\kern 1pt & 0\kern 1pt }, \quad \delta_3=\pmatrix{0\kern 1pt &0\kern 1pt \cr
1\kern 1pt & 0\kern 1pt }, \quad \delta_4=\pmatrix{0\kern 1pt &0\kern 1pt \cr
0\kern 1pt & 1\kern 1pt }\leqno (1.2)$$
So the family of block-diagonal $2N\times2N$ matrices with $2\times2$ entry $\delta_j$, $1\leq j\leq4$ at the $i$:th place, $1\leq i\leq N$
$$\bigl(D^i_j\bigr)_{i\in\Lambda\cup\partial\Lambda}=\bigl(0\oplus\cdots\oplus\delta_j\oplus\cdots\oplus0
\bigr)\leqno(1.3)$$
gives an orthonormal basis (ONB) of 1-point functions $\widetilde{\cal O}_{\bf R}$.
\smallskip
\noindent {\it Example 2}:
$\widetilde{\cal O}_{\bf R}$ is the algebra generated by Pauli matrices $(\widetilde D^i)_{i\in\Lambda\cup\partial\Lambda}$ with
diagonal block supported on site $i$ that takes values in
$\{\Id,i\sigma^x,i\sigma^y,i\sigma^z\}$. 
\smallskip
We shall restrict to the canonical algebra, whose generators enjoy the nice property of being real 
matrices.
Let also $o_{\bf R}\subset o$ be the algebra of $2\times 2$ matrices with real coefficients,
endowed with the scalar product $(A|B)=\Tra (B^*A)$, which is isometric with ${\bf R}^4$.
By extension, the basis $\delta=\{\delta_1,\delta_2,\delta_3,\delta_4\}$ of $o_{\bf R}$ will be called
an ``elementary basis'' of $\widetilde{\cal O}_{\bf R}$, since
$N$ copies of $\delta$, attached to each site $i$, give a basis $(D^i_j)_{i\in\Lambda\cup\partial\Lambda,1\leq j\leq4}$ of $\widetilde{\cal O}_{\bf R}$.
We say the same thing of any other ONB $b=\{b_1,b_2,b_3,b_4\}$ of $o_{\bf R}$, and of the corresponding basis
$(B^i_j)_{i\in\Lambda,1\leq j\leq4}$ of $\widetilde{\cal O}_{\bf R}$, where $B^i_j$ is defined as in (1.3), with $b_j$
instead of $\delta_j$.
Actually, the order of the elements of $b$ matters, so we prefer the matrix notation, namely
$$b=\pmatrix{b_1\kern 1pt &b_2\kern 1pt \cr
b_3\kern 1pt & b_4\kern 1pt }\in{\cal M}_{4\times4}({\bf R})\leqno(1.4)$$
is a $2\times2$ block-matrix, where $b_k$ is of the form
$$b_k=\pmatrix{b^{1k}\kern 1pt &b^{2k}\kern 1pt \cr
b^{3k}\kern 1pt & b^{4k}\kern 1pt }\in{\cal M}_{2\times2}({\bf R})\leqno(1.5)$$
which we identify with the vector $b^k={}^t \bigl(b^{1k},b^{2k},b^{3k},b^{4k}\bigr)$.
Actually we will never use the algebraic structure of $o_{\bf R}$ in this paper.

Consider now $b$ as a linear operator on ${\cal K}_1\otimes{\cal K}_2$, with ${\cal K}_1={\cal K}_2={\bf R}^2$, and recall [Si,Sect.II.1]
the partial trace $\tra _1$ from ${\cal L}({\cal K}_1\otimes{\cal K}_2)$ to ${\cal L}({\cal K}_1)$ is defined by the
requirement~:
$$\tra _{{\cal K}_1}\bigl(X({\tra}_1(b))\bigr)= {\tra}_{{\cal K}_1\otimes{\cal K}_2}((X\otimes1)b), \quad X\in{\cal L}({\cal K}_1)$$
where $\tra={1\over d}\Tra$, $d$ is the dimension and Tr the ordinary trace. Then with the notations above, we see easily that~:
$$\tra _1(b)=\pmatrix{\tra b_1&\tra b_2\cr\tra b_3&\tra b_4\cr}\leqno(1.6)$$
So $\tra_1$ are the components, in some matrix representation, of the usual trace, or (``tracial state'') on $o_{\bf R}$.
For simplicity we set $T(b)=\tra _1(b)$ and call it the ``matrix of traces''.
An important r\^ole will be played with symmetric basis. 
\medskip
\noindent {\bf Definition 1.1}: 
{\it We call the ONB $b$ {\it symmetric} iff the corresponding matrix $b$ in (1.4) is Hermitian, i.e. $b_1=b_1^*$,
$b_4=b_4^*$, and $b_3=b_2^*$. We call it $\delta$-{\it symmetric} if moreover $b$ is real, and
$T(b)$ has a degenerate eigenvalue, that is, is a multiple of identity. We denote by $[\delta]_s$ the class of $\delta$-symmetric basis.}
\smallskip
Most of the basis are not symmetric, but occasionally we can make them symmetric,
by permuting or multiplying by $-1$ some elements. 
Note that $\{\Id,i\sigma^x,i\sigma^y,i\sigma^z\}$ cannot be made symmetric; this is one of the reasons why we prefer
the canonical basis $\{\delta_1,\delta_2,\delta_3,\delta_4\}$. In Appendix, we characterize $\delta$-symmetric basis, up to 
such transformation.

So far we have constructed ``one point functions'', i.e. a basis of $\widetilde{\cal O}_{\bf R}$. In the sequel we content
with Hamiltonians of type (0.1) which are of 
second order in the interactions; if we were to include the linear term
$\Sum_{i\in\Lambda}\sigma_i^z$ we would write it as $\Sum_{\langle i,j\rangle}1_i\otimes\sigma_j^z$.
Embed $\widetilde{\cal O}_{\bf R}$ into
$\widetilde{\cal O}_{\bf R}\otimes\widetilde{\cal O}_{\bf R}$ by the usual coproduct $\Delta$, and set
$\widetilde x=\Delta(x)={1\over2}(1\otimes x+x\otimes1)\in o_{\bf R}\otimes o_{\bf R}$, for $x\in o_{\bf R}$.
So we have ``lifted'' $\widetilde b=\Delta(b)$ as a family of $\widetilde{\cal O}_{\bf R}\otimes\widetilde{\cal O}_{\bf R}$
by $(\widetilde B^i_j)_{i\in\Lambda,1\leq j\leq4}$, with $\widetilde B^i_j=\Delta(B^i_j)$.
With the notations of (1.4) and (1.6) we have
$$\widetilde B^i=\pmatrix{\widetilde B^i_1\kern 1pt &\widetilde B^i_2\kern 1pt \cr
\widetilde B^i_3\kern 1pt & \widetilde B^i_4\kern 1pt }\in{\cal M}_{4N\times4N}({\bf R}), \quad
\tra _1(\widetilde B^i)=\pmatrix{\tra \widetilde B^i_1\kern 1pt &\tra \widetilde B^i_2\kern 1pt \cr
\tra \widetilde B^i_3\kern 1pt & \tra \widetilde B^i_4\kern 1pt }\in{\cal M}_{2\times2}({\bf R})
\leqno(1.8)$$
In the same way, we form $e^{-\beta H}\widetilde B^i_j$,
so we can map to each site $i\in\Lambda\cup\partial\Lambda$ a $2\times2$ matrix~:
$$\tra _1(e^{-\beta H}\widetilde B^i)=\tra (e^{-\beta H})\pmatrix{\omega_\beta(\widetilde B^i_1)\kern 1pt
&\omega_\beta(\widetilde B^i_2)\kern 1pt \cr
\omega_\beta(\widetilde B^i_3)\kern 1pt & \omega_\beta(\widetilde B^i_4)\kern 1pt }\leqno(1.9)$$
\medskip
\noindent{\bf Definition 1.2}:
{\it We call {\it vorticity matrix} at site $i$, relative to the basis $b$, at inverse temperature $\beta$, the matrix~:
$$\Omega^i_\beta(b)={\tra _1(e^{-\beta H}\widetilde B^i)\over\tra (e^{-\beta H})}$$
The traceless matrix $$\widehat \Omega^i_\beta(b)=\Omega^i_\beta(b)-
\tra \bigl(\Omega^i_\beta(b)\bigr)\Id\leqno(1.10)$$
is called  the {\it reduced vorticity matrix} at site $i$.}
\medskip
\noindent {\it Example}: $\Lambda=\{1,2\}$ is a lattice with 2 sites, $\partial\Lambda=\emptyset$, one has
$\widehat \Omega^1_\beta(\delta)=\widehat \Omega^2_\beta(\delta)=0$. This is observed also numerically
for all $\Lambda$, with $\partial\Lambda=\emptyset$, although vortices should merge sponteanously in infinite volume.
\smallskip
If $b$ is a symmetric basis of $o_{\bf R}$, then $\Omega^i_\beta(b)$ and $\widehat\Omega^i_\beta(b)$
are hermitean since $H$ is self-adjoint (and are real symmetric if $H$ moreover has real coefficients), and 
$$\bigl(\widehat \Omega^i_\beta(b)\bigr)^2=\det\widehat \Omega^i_\beta(b)\Id\leqno(1.11)$$
Thus $\Omega^i_\beta(b)$ enjoys the nice property, to be diagonalizable with real (opposite) eigenvalues for all sites $i$,
and all inverse temperature $\beta$. Viewing these as a field of matrices over the lattice, we can figure out
the ``vorticity'' of the system (whenever this makes sense), by simply looking at their principal directions.
This also gives a measure of vorticity, i.e. numbers (integers) that should be independent of the choice of ``elementary'' basis
$b$.

Next we define vortices as the set of sites where the reduced vorticity matrix is singular.
\medskip
\noindent {\bf Definition 1.3}: {\it We say that $\xi\in\Lambda$ is a {\it vortex} at inverse temperature $\beta$
relative to the $\delta$-symmetric ONB $b$ iff
$\Omega^\xi_\beta(b)$ has a degenerate eigenvalue, i.e. $\widehat\Omega^\xi_\beta(b)=0$.
We call {\it regular} the other points.}
\smallskip
By construction, all sites are vortices when $\beta=0$.
\smallskip
Now we turn to consistency of Definitions 1.2 and 1.3 relatively to the choice of $b$ within $\delta$-symmetric basis.
That $b$ is a $\delta$-symmetric basis is a natural requirement for computing the degree of $\widehat\Omega^i(b)$, see Sect.2.
With $b$ written as in (1.4), and  $P\in O(2;{\bf R})$, we set with obvious notations 
$$a={}^tPbP\leqno(1.12)$$ 
(i.e. as if $b_j$'s were numbers). 
The same holds after taking the co-product $\Delta$ of each term, i.e. $\widetilde a={}^tP\, \widetilde b P$.
This defines conjugacy classes, which pass to the partial traces (1.6), i.e. 
$T(a)={}^tPT(b)P$, and $T(\widetilde a)={}^tP\, T(\widetilde b)P$.
Moreover, if $X\in{\cal L}({\bf R}^2)$, we have 
$$(1\otimes X)b=\pmatrix{Xb_1\kern 1pt &Xb_2\kern 1pt \cr Xb_3\kern 1pt &Xb_4\kern 1pt}=(1\otimes X)Pa\, {}^tP\leqno(1.13)$$  
After lifting $\widetilde a$ and $\widetilde b$ to $\widetilde{\cal O}_{\bf R}\otimes\widetilde{\cal O}_{\bf R}$, 
(1.8) becomes
$$\widetilde A^i={}^tP\pmatrix{\widetilde B^i_1\kern 1pt &\widetilde B^i_2\kern 1pt \cr
\widetilde B^i_3\kern 1pt & \widetilde B^i_4\kern 1pt }P\in{\cal M}_{4N\times4N}({\bf R}), \quad
\tra _1(\widetilde A^i)={}^tP\pmatrix{\tra \widetilde B^i_1\kern 1pt &\tra \widetilde B^i_2\kern 1pt \cr
\tra \widetilde B^i_3\kern 1pt & \tra \widetilde B^i_4\kern 1pt }P\in{\cal M}_{2\times2}({\bf R})
\leqno(1.14)$$
and (1.13) also extends when taking $X\in{\cal L}({\bf R}^{4N})$ and replacing $a$ by $\widetilde A^i$, $b$ by $\widetilde B^i$. 
Let now $X=e^{-\beta H}$, we obtain that conjugacy classes pass to vorticity matrices, i.e. 
$$\Omega^i_\beta(a)={}^tP\Omega^i_\beta(b)P,\quad \widehat\Omega^i_\beta(a)={}^tP\widehat\Omega^i_\beta(b)P\leqno(1.15)$$
In Appendix we characterize completely the set of $\delta$-symmetric ONB's
up to the action of the group $G_0$ acting on ${\cal M}_4({\bf R})$ by 
permutations, or multiplication by $-1$, of some vectors $b^k\in{\bf R}^4$ (recall from (1.5)
the identification of $b^k$ with the $2\times2$ matrix $b_k$). Namely, we show in Appendix that if $b$ is $\delta$-symmetric,
then modulo the action of $G_0$, 
there exists discrete or one-parameter families $P_s\in O(2;{\bf R})$ such that 
$$b=b_s={}^tP_s\delta P_s\leqno(1.16)$$
So we have~:
\medskip
\noindent {\bf Proposition 1.4}: {\it Definitions 1.2 and 1.3 are consistent, i.e. vorticity matrices relative to all
$\delta$-symmetric ONB $b$ are related by (1.15) for some $P_s\in O(2;{\bf R})$, and in particular
$\xi$ is a vortex relative to $\delta$ iff this is a vortex relatively to any $\delta$-symmetric $b$.}
\smallskip
Moreover we have the numerical evidence that, among all $\delta$-symmetric basis $b$, 
the canonical basis $\delta$ is most ``faithful'', in the sense that 
$\Omega^i_\beta(\delta)$ have on the boundary lattice $\partial\Lambda$ the same principal directions as the directions along which
Pauli matrices are compressed (associated with the eigenprojector $\Pi_j$). 

Let us conclude this section by some heuristic remark:
In the case of a translation invariant interaction
$\Phi(X)$, and free boundary condition ($\partial\Lambda=\emptyset$) it is standard to show that Gibbs state (0.1) converges
in the thermodynamical limit, and so do the vorticity matrices. This is already of interest, because in the framework of  
Mermin-Wagner and Hohenberg theorems, it is believed that (quantum) vortices merge spontaneously in infinite volume, 
without any boundary conditions. In our case however, the compression of Pauli matrices on $\partial\Lambda\neq\emptyset$
breaks the translation invariance. When $\partial\Lambda={\bf Z}^2\setminus\Lambda$, we could think of the boundary condition as an
external field, that we recall from [Si,Sect.II 3]~:

A state is the assignment of an 
operator $\rho_X$ for each finite $X\subset{\bf Z}^2$ with $\Tra (\rho_X)=1$ and 
$\tra _{{\cal H}_Y}(\rho_{X\cup Y})=\rho_X$ (partial trace
on ${\cal H}_Y=\otimes_{i\in Y}{\bf C}^2_i$,~) for all disjoint $X,Y\subset{\bf Z}^2$. 
Given the nearest neighbor interaction $\Phi(X)$, and a state $\rho$ on ${\bf Z}^2$, we define the Hamiltonian on all ${\bf Z}^2$
$$H^\rho_\Lambda(\Phi)=-\Sum_{X\cap\Lambda\neq\emptyset}\Tra _{X\setminus\Lambda}
\bigl[(1\otimes\rho_{X\setminus\Lambda})\Phi(X)\bigr]$$
that couples $\Lambda$ with the external field $\rho$ through its nearest neighbors at the boundary. 

For any quasi-local observable $A$ on ${\bf Z}^2$, define the expectation value
$$\langle A\rangle^\rho_{\beta,\Lambda}={\Tra _\Lambda\bigl(\exp\bigl[-\beta H^\rho_\Lambda(\Phi)\bigr]A\bigr)\over 
\Tra _\Lambda\exp\bigl[-\beta H^\rho_\Lambda(\Phi)\bigr]}$$
Then it is known that both 
$|\Lambda|^{-1}\log \Tra _\Lambda\exp\bigl[-\beta H^\rho_\Lambda(\Phi)\bigr]$ and 
$|\Lambda|^{-1}\log \Tra _\Lambda\bigl(\exp\bigl[-\beta H^\rho_\Lambda(\Phi)\bigr]A\bigr)$ have a limit as $|\Lambda|\to\infty$, 
and so has $\langle A\rangle^\rho_{\beta,\Lambda}$. On that basis we could expect that the vorticity matrices 
with a conditional external field have a limit in the thermodynamical limit.
However, it is not quite clear which field $\rho$ could stand for the compression of Pauli matrices at the boundary
(see [AscPil] for the 1-D XY chain);
our Hamiltonian $H_{\Lambda\cup\partial\Lambda}(\Phi)$ is only an approximation for $H^\rho_\Lambda(\Phi)$. 


\bigskip
\noindent {\bf 2. Holonomy on the Lie group $\Ssl(2;{\bf R})$}. 
\smallskip
We pass here to an idealistic continuous limit, where vorticity matrices would be defined as a smooth field on ${\bf R}^2$
(away from vortices) , 
valued in the Lie algebra $\ssl (2;{\bf R})$, consisting of traceless matrices. Our purpose is to integrate such fields vanishing at some points,
and define the ``non-commutative degree''.
For advanced results on Differential Calculus 
on lattices in the scalar case, see [Sm]. The non-commutative discrete case, also allowing for an extension of our XY model 
to Heisenberg model, has still to be set up. 

Let $M:D\subset{\bf R}^2\to\ssl (2;{\bf R}), x\mapsto M(x)$ be a $C^1$ map, 
such that such that $M(x)$ obeys (1.11), i.e. $M(x)^2=\lambda(x)\Id$, $\lambda(x)\geq0$, and consider 
$\rho\in\Lambda^1({\bf R}^2;\ssl (2;{\bf R}))$ the 1-form defined by
$$\rho(x)={1\over 2}(M^{-1}(x)dM(x)-dM(x)M^{-1}(x))\leqno(2.1)$$
with the property of being antisymmetric if $M$ is symmetric.
Since 
$$M^{-1}(x)dM(x) +dM(x)M^{-1}(x))={d\lambda(x)\over\lambda(x)}$$ 
we have
$$M^{-1}(x)dM(x)=\rho(x)+{d\lambda(x)\over2\lambda(x)}, \quad dM(x)M^{-1}(x)=-\rho(x)+{d\lambda(x)\over2\lambda(x)}$$
If moreover $D\subset{\bf R}^2$ is simply connected and $\lambda(x)>0$ in $D$, 
${d\lambda(x)\over2\lambda(x)}$ is exact, and if $\gamma$ is a loop in $D$~: 
$$\int_\gamma M^{-1}(x)dM(x)=-\int_\gamma dM(x)M^{-1}(x)=\int_\gamma\rho(x)\leqno(2.3)$$
In the general case, using identity $M(x)^2=\lambda(x)\Id $, we find easily
$$
d\rho(x)=\bigl[M{\partial M\over\partial x_1}M{\partial M\over\partial x_2}-
M{\partial M\over\partial x_2}M{\partial M\over\partial x_1}-{\partial M\over\partial x_2}M{\partial M\over\partial x_1}M+
{\partial M\over\partial x_1}M{\partial M\over\partial x_2}M
\bigr]{dx_1\wedge dx_2\over\lambda^2(x)}
$$
and setting  $M=\pmatrix{a\kern 1pt &b\kern 1pt \cr
c\kern 1pt & -a\kern 1pt }$, a computation shows that
$$d\rho=-\lambda^{-2}(a db\wedge dc+b dc\wedge da+ c da\wedge
db)M\leqno(2.5)$$
Thus the form $\rho$ is closed  if
$$R(a,b,c)=a db\wedge dc+b dc\wedge da+ c da\wedge db=0\leqno(2.6)$$ 
and this condition holds if $M$ is symmetric.
For such a map $M$ uniformly elliptic at infinity, in the sense that
$$|\lambda(x)|\geq C>0, \quad |x|\geq r_0$$
we can define the number
$$s_\infty=\det {1\over 2\pi}\int_{|x|=r}\rho(x)\leqno(2.7)$$
which,  by Stokes' formula, turns out to be independent of $r\geq r_0$. In the same way, if 
$\xi$ is a vortex (i.e. the map $M$ is singular at $\xi$) we define the  ``local degre local''
$$s_\xi=\det {1\over 2\pi}\int_{\gamma}\rho(x) \leqno(2.8)$$ 
whenever $M(x)$ is invertible
for $x\neq \xi$, integrating on a small contour $\gamma$ around $\xi$. 

Let us now compute Maurer-Cartan structure equation for the form $\rho(x)$
[Ma,p.165]. The structure coefficients for the Lie algebra $\ssl (2;{\bf R})$, with basis 
$e_1=\pmatrix{1&0\cr 0&-1}, e_2=\pmatrix{0&1\cr 0&0}, e_3=\pmatrix{0&0\cr 1&0}$
are given by $C^2_{1,2}=-C^2_{2,1}=2,C^3_{1,3}=-C^3_{3,1}=-2,C^1_{2,3}=-C^1_{3,2}=1$, and
$C^k_{i,j}=0$ otherwise. For $[\rho,\rho]=
\Sum_k(\Sum_{i<j}C^k_{i,j}\rho^i\wedge\rho^j)e_k$, we find
$$-[\rho,\rho]=\lambda^{-2}(a db\wedge dc+b dc\wedge da+ c da\wedge db)M$$
Under (2.6), this relation together with (2.5) show that $\rho$ verifies
$$d\rho+[\rho,\rho]=0\leqno(2.9)$$ 
Recall that if $G$ is a Lie group, 
and ${\cal A}$ its Lie algebra, $\omega$ the canonical 
Maurer-Cartan form on $G$, invariant by left translations, we 
define Darboux differential of the map $f\in C^1(D;G)$
by $\pi_f=f^*\omega$. The fundamental existence theorem (``Poincar\'e lemma''),
with a differential form
$\rho\in\Lambda^1(D;{\cal A})$ verifying $d\rho+[\rho,\rho]=0$, associates (locally) a map $f\in C^1(D;G)$,  
whose Darboux differential is precisely equal to $\rho$. Moreover this map is unique
when assigning its value on a point $x_0\in D$.

On the other hand we know [Ki,p.117\& 321], that the Lie group whose Lie algebra is $\ssl (2;{\bf R})$,
is the universal covering $E=\widetilde{\Ssl}(2;{\bf R})$ of the unimodular group
$\Ssl (2;{\bf R})$. The unimodular group is 
topologically equivalent to the cylinder ${\bf S}^1\times{\bf R}^2$, its  
fundamental group equals ${\bf Z}$, and  
$E$ is homeomorphic to ${\bf R}^3$. (It is known however that one cannot parametrize $E$ by matrices, more precisely
$E$ cannot be written as a subgroup of some GL$(m;{\bf C})$, with $m\in{\bf N}$, but rather as a tensor product of such matrices.~)

Relation (2.9) ensures the existence of a local primitive $N\in C^1(D;\widetilde{\Ssl}(2;{\bf R}))$ of $\rho$,
the ``logarithm'' of $M$. If $D$ is simply connected, this primitive is also global. Otherwise, consider its extension to $E$, and 
let $\gamma\subset{\bf R}^2$ be a loop at $x_0\in D$, we may define the monodromy opeerator $T_\gamma$  acting on functions
$N:\gamma\to E$. 
\medskip
\noindent{\it Example 1}: For symmetric matrices in $\ssl (2;{\bf R})$
$$M_0(x)=\pmatrix{\cos n\theta &\sin n\theta \cr\sin n\theta & -\cos n\theta },\quad
M_r(x)=\pmatrix{r\cos n\theta &\sin n\theta \cr \sin n\theta  & -r\cos n\theta }\leqno(2.10)$$
we have $(\deg _\infty(M))^2=n^2$. The 1-form $\rho$ associated wih $M_0$ is simply 
$\pmatrix{0&n \cr -n& 0\kern 1pt }d\theta$. 
\medskip
Since the fundamental group of $E$ is ${\bf Z}$, $s_\infty$ and $s_\xi$ are integers so that we set 
$$s_\infty=(\deg _\infty(M))^2\in\{0,1,4,9,\cdots\}, \quad s_\xi=(\deg _{x_1}(M))^2\in\{0,1,4,9,\cdots\}\leqno(2.11)$$
Degrees at infinity and at $x_0$ are then  invariant by homotopy. In particular, a perturbation theory can be carried out by
expanding $M$ as Fourier series.
The degree for matrices $M\in\ssl (2;{\bf R})$ verifiant (2.6) can be also obtained by Brouwer theory
[Mi], by considering these matrices as  (locally) a 2-D manifold.
If  matrices $M$ are symmetric, then $\rho$ is antisymmetric, and 
$${1\over 2\pi}\int_\gamma\rho=\pmatrix{0&-n\cr n&0\cr}\leqno(2.12)$$
with $n\in{\bf Z}$. 


\bigskip
\noindent{\bf 3. Numerical simulations}.
\medskip
Recall we have completed the lattice $\Lambda$ with an environment $\partial\Lambda\subset{\bf Z}^2$
where Pauli matrices are compressed in directions $(\theta_j)_{j\in\partial\Lambda}$, i.e. we change $\sigma$ by
$\Pi_\theta\sigma\Pi_\theta$, where
$\Pi_\theta=\pmatrix{\cos^2{\theta} &\cos{\theta}\sin{\theta} \cr \cos{\theta}\sin{\theta} &\sin^2{\theta}}$
Thus
$$\sigma^x_i(\theta_i)=(\sin 2\theta_i)\Pi_{\theta_i}, \quad \sigma^y_i(\theta_i)=0\leqno(3.1)$$
Hamiltonian (0.1) with nearest neighbor interaction has a large kernel, so it is not directly suitable for
numerical simulations, even when modified by an external field. In the Classical case,
we consider instead the 2-D planar rotator with long range interactions, such as Kac's model [El-BoRo]; 
renormalizing the Hamiltonian leads to the free
energy functional on macroscopic scales (or coarse graining), whose critical points are most accessible to numerical analysis. 
But in the Quantum case, renormalization procedures for 2-D planar rotator with long range interactions are not yet available
(see however e.g. [ScOr] for Ising model). 
Changing (0.1) to the anisotropic XY model is in fact a first attempt to lift the degeneracy of the spectrum 
of the Hamiltonian, and enhance the effects of the external field on vorticity
even in quite small lattices, so that the predicted vorticity could be observed with a fairly good accuracy, only using elementary 
numerical tools (Wolfram Mathematica on a laptop). We discuss below the r\^ole of anisotropy.
For $n,k>0$, consider the Hamiltonian
$$\eqalign{
&H_{(n,k)}(\sigma|\partial\Lambda) =
-{1\over2(n+k)} \Sum_{\langle i,j\rangle ; i,j\in\Lambda}
({n}\, \sigma_i^x\otimes\sigma_j^x +
{k}\, \sigma_i^y\otimes\sigma_j^y)\cr
&-{1\over2(n+k)} \Sum_{\langle i,j\rangle ; (i,j)\in\Lambda\times\partial\Lambda}
{n}\, (\sigma_i^x\otimes\sigma_j^x(\theta_j)+
\sigma_j^x(\theta_j)\otimes\sigma_i^x)
-{1\over2(n+k)} \Sum_{\langle i,j\rangle ; i,j\in\partial\Lambda}
{n}\, \sigma_i^x(\theta_i)\otimes\sigma_j^x(\theta_j)\cr
}\leqno(3.2)$$
so $H_{(n,k)}(\sigma|\partial\Lambda)$ is self-adjoint and real.
When $n=k=1$, $\Lambda={\bf Z}^2$, $H=H_{(1,1)}$ is the most natural (isotropic) model with $O^+(2)$ symmetry.
It enjoys nice properties, like reflection positivity;
its spectrum is believed to be absolutely continuous on $[-2,2]$
as this of the Laplacian on ${\bf Z}^2$, but this is not rigorously known, see [DaManTie], [De].
For $k\neq1$, we call $H_{(1,k)}(\sigma|\partial\Lambda)$  the {\it anisotropic XY model}. Only when $\partial\Lambda=\emptyset$,
$H_{(1,k)}$ is unitarily equivalent to $H_{(k,1)}$.
In general, $H_{(1,k)}(\sigma|\partial\Lambda)$ has no obvious symmetry,
but it is most suitable for studying vorticity matrices on finite lattices $\Lambda\cup\partial\Lambda$,
at least for small $\beta$.

We consider rectangular lattices of minimal sizes to exclude important volume effects, with sufficiently large $\partial\Lambda$
to constrain the ``quantum system'' within $\Lambda$. We choose $\theta_j=d\omega_j+\phi$ where $\omega_j$ is the polar angle
representing the vector $j\in\partial\Lambda$. We compare calculations for $n=1$ and $k=1,2,10$. 
\medskip
\noindent {\it a) Considerations on spectra}.
\smallskip
Numerically, we observe that the spectrum of $H_{(1,k)}$ is distributed
in an interval $I$ close to $[-2,2]$,
and looks symmetric around 0, allowing for equivalence between ferromagnetic Hamiltonian $H_{(1,k)}(\sigma|\partial\Lambda)$
and antiferromagnetic $-H_{(1,k)}(\sigma|\partial\Lambda)$.
The distribution has smaller density
at the edges of $I$, and larger near $\lambda=0$.
We present below the integrated density (statistical distribution) of states
$\rho(\lambda) = \#\{\lambda_k | \lambda_k < \lambda\}$
in various situations,
namely we compare isotropic and anisotropic cases
without boundary (Fig.1 a, b)
and anisotropic case with boundary (Fig.1 c).
\bigskip
\input epsf \centerline{(a)\epsfxsize=7truecm \epsfbox{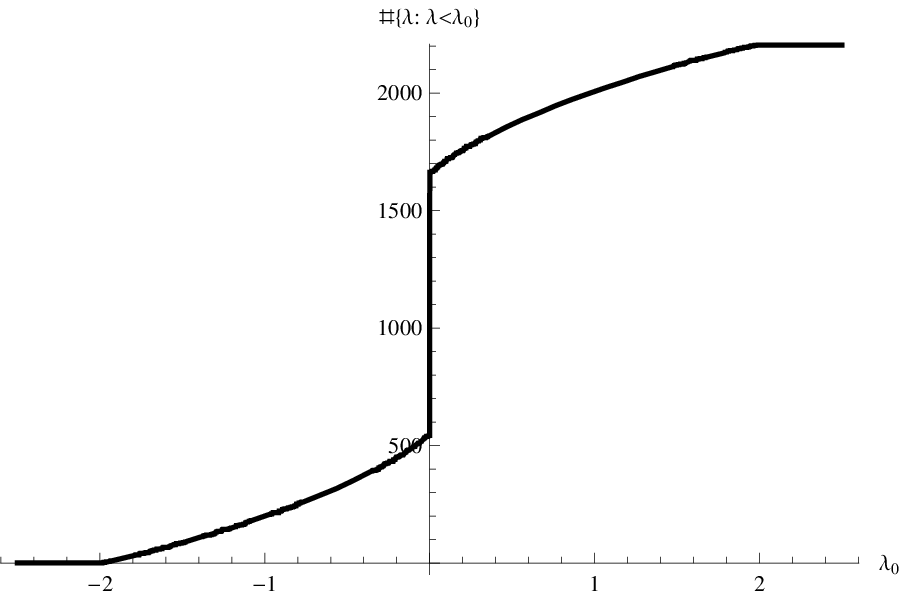}}
\medskip
\input epsf \centerline{(b)\epsfxsize=7truecm \epsfbox{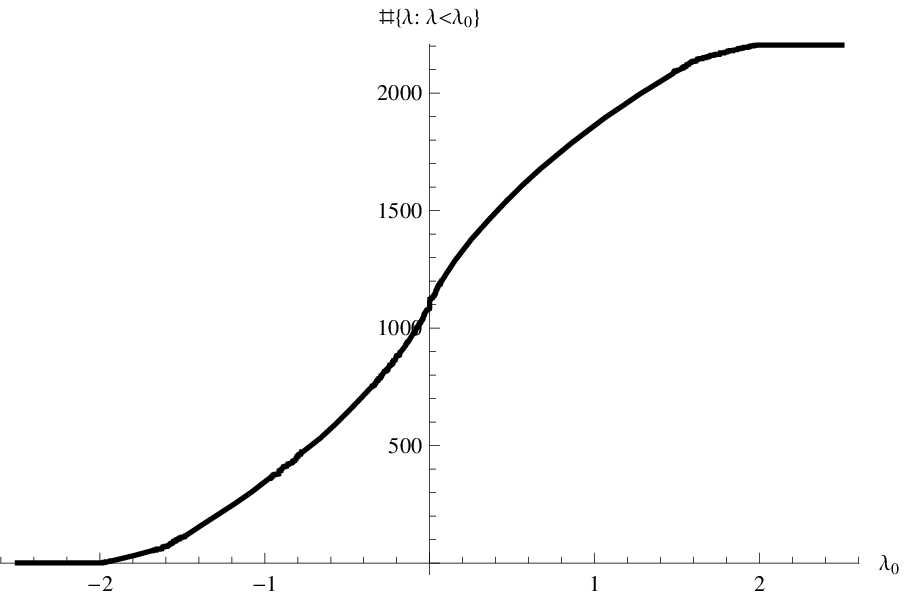}}
\medskip
\input epsf \centerline{(c)\epsfxsize=7truecm \epsfbox{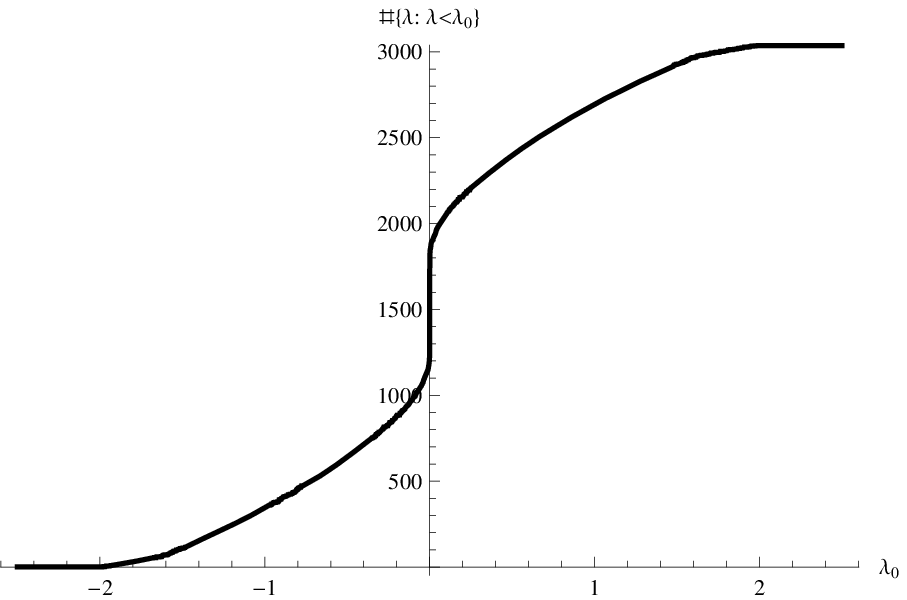}}
\medskip
{\it Fig.1.} {The integrated density of states
$\rho(\lambda) = \#\{\lambda_k | \lambda_k < \lambda\}$:
(a) isotropic case $H_{(1,1)}$ without boundary, $|\Lambda|=19\times 29$;
(b) anisotropic case $H_{(1,10)}$ without boundary, $|\Lambda|=19\times 29$;
(c) anisotropic case $H_{(1,10)}$ with 2 boundary layers,
$|\Lambda\cup\partial \Lambda| = 23\times 33$.}
\medskip
The reason for degeneracy at $\lambda = 0$ is the following.
In the isotropic case $n=k=1$ the matrix $\sigma_i^x\otimes\sigma_j^x+\sigma_i^y\otimes\sigma_j^y$ is of rank 2,
so when $\partial\Lambda=\emptyset$, half of the eigenvalues of $H$ vanish, and also in the general case there is a big degeneracy
of the spectrum near $\lambda=0$.
As in QFT we could try to remove that
``artificial'' part of $\ker H$ by reducing the Hilbert space ${\cal H}={\bf C}^{4N}$ to a ``physical space'', but a difficulty
arises because $H$ is not positive in the form sense.
The effect of anisotropy is to lift this degeneracy, and enhance vorticity effects.
Another reason for degeneracy is boundary effects as can be seen from comparison between Fig.~1.b and 1.c.
Degeneracy could be also reduced by enlarging the inner lattice, but at the expense of computational difficulties.
\medskip
\noindent {\it b) Vorticity patterns}.
\smallskip
We study Gibbs state at inverse temperature $\beta$, with significant results provided $\beta$ ranges in some interval, for which
however, there is no evidence of a second order phase transition.

To visualize monodromy of the vorticity matrices $\Omega^i_\beta(\delta)$, we plot their principal directions as ``crosses'', of length
proportional to their eigenvalues (recall the reduced vorticity matrices are symmetric, with eigenvalues $\pm\sqrt{\lambda_i}$,~)
as we would do with arrows in the classical model [El-BRo].
As expected, their principal directions coincide on $\partial\Lambda$, with those of the eigenprojectors $\Pi_{\theta_i}$.
We expect also the number of vortices to be equal to the topological degree $d=s_\infty$.
Indeed, computing ${1\over 2\pi}\int_\gamma\rho$
as a discrete integral along a contour $\gamma\in\Lambda$,
not too far from the boundary (in practice, 2 or 3 layers),
it turns out that the computed degree is close
to this we would obtain in example (2.12). 

Because of degeneracy in the isotropic case 
vorticity matrices in all sites that are not the first neighbors to the boundary
are of the form
$\Omega_\beta^i(\delta)\approx\pmatrix{c^i(\beta) & 0 \cr 0 & c^i(\beta)}$ for some $c^i(\beta)$,
and so the reduced vorticity matrices are zero $\tilde \Omega_\beta^i(\delta) = 0$ in these sites.
This fact can be shown by direct calculations of eigenvectors and vorticity matrices
and is illustrated in Fig.~2 below.
\bigskip
\input epsf \centerline{(a)\epsfxsize=7truecm \epsfbox{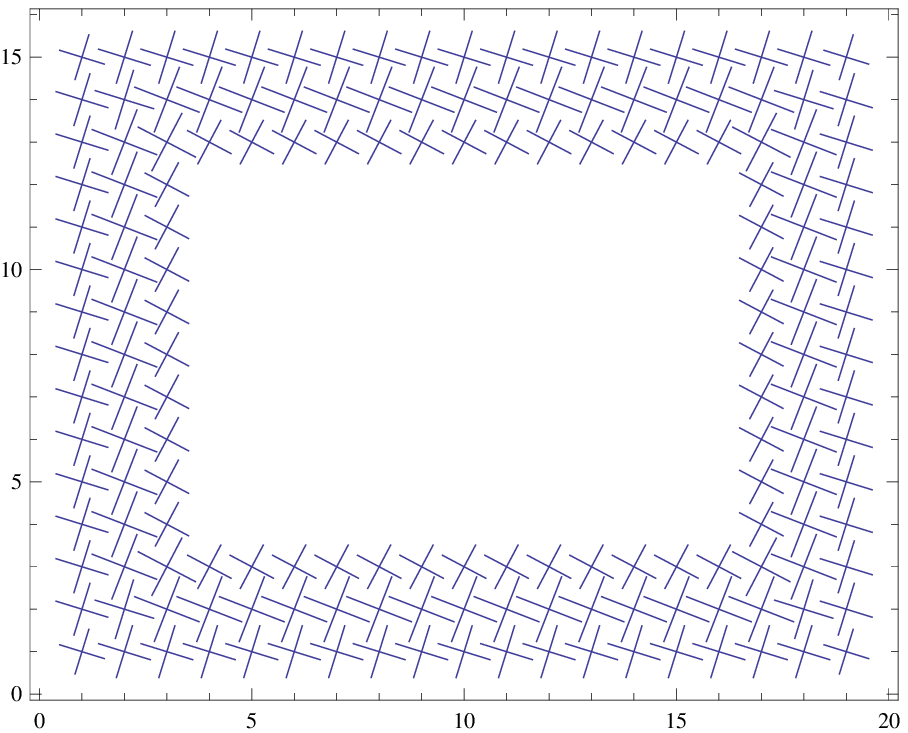}
(b)\epsfxsize=7truecm \epsfbox{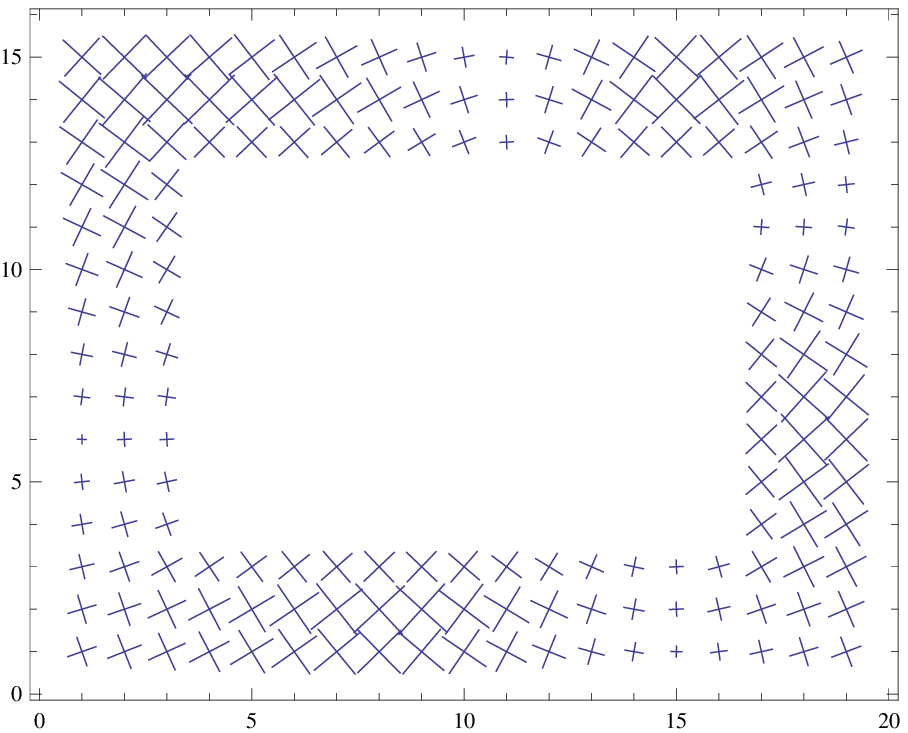}}
\medskip
{\it Fig 2.} {Principal directions (lengths correspond to sizes of eigenvalues on a logarithmic scale) for different degree: (a) $d=0$, (b) $d=1$.
Lattice is $|\partial \Lambda \cup \Lambda| = 15\times19$
with 2 boundary layers.
Here is considered the isotropic Hamiltonian $H_{(1,1)}$ with $\beta=1$.}
\medskip
If we consider anisotropic case then matrices $\sigma_i^x\otimes\sigma_j^x+\sigma_i^y\otimes\sigma_j^y$
are no longer degenerate and so eigenvalues in the center are no more zeros
(see Fig.~3).
In this case eigenvalues decay
when getting far from the boundary towards the center of $\Lambda$
and it seems that the rate of decay is exponential
with distance from the boundary.
To make the vorticity patterns more demonstrative
we draw the crosses on a logarithmic scale.
All points sufficiently close to the center of $\Lambda$
look like vortices within the standard accuracy of computations,
but the number of ``true'' vortices should be equal
to the topological degree $d=s_\infty$.
\bigskip
\input epsf \centerline{(a)\epsfxsize=7truecm \epsfbox{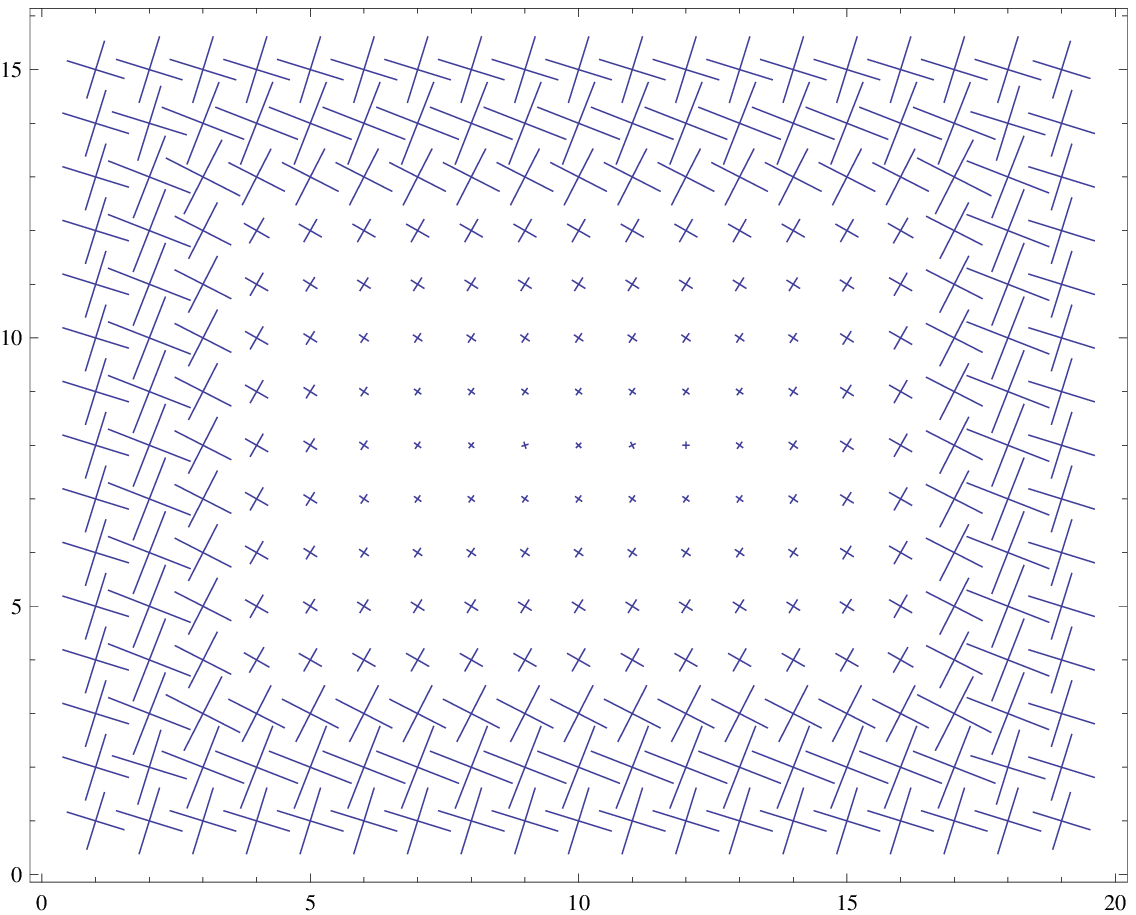}
(b)\epsfxsize=7truecm \epsfbox{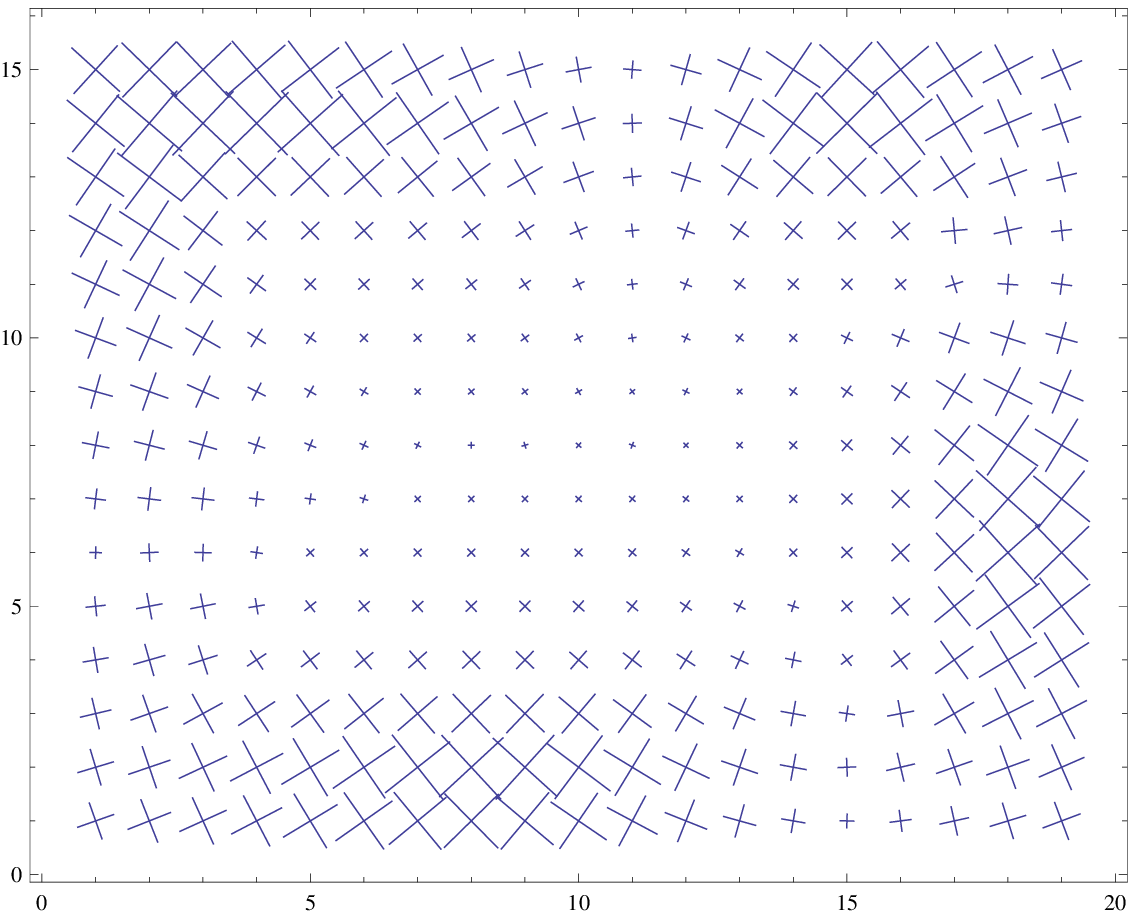}}
\input epsf \centerline{(c)\epsfxsize=7truecm \epsfbox{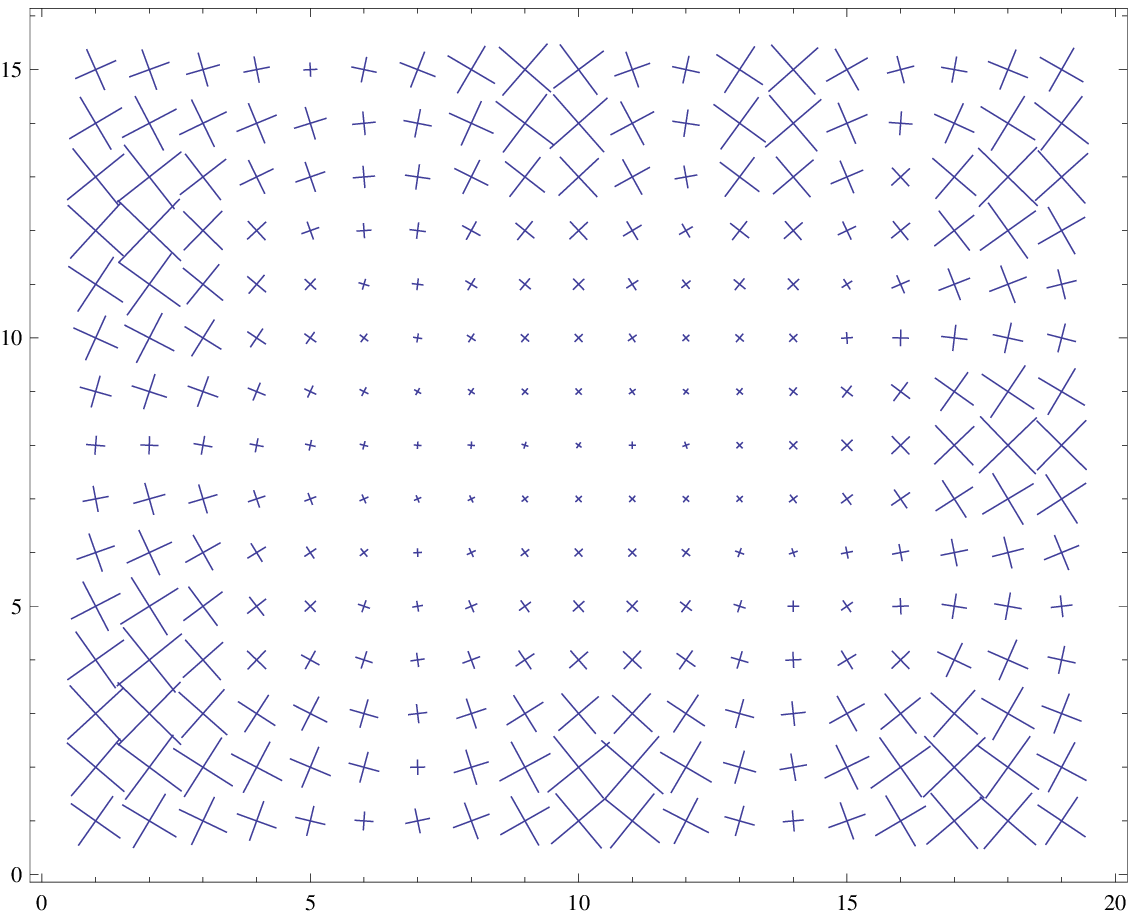}
(d) \epsfxsize=7truecm \epsfbox{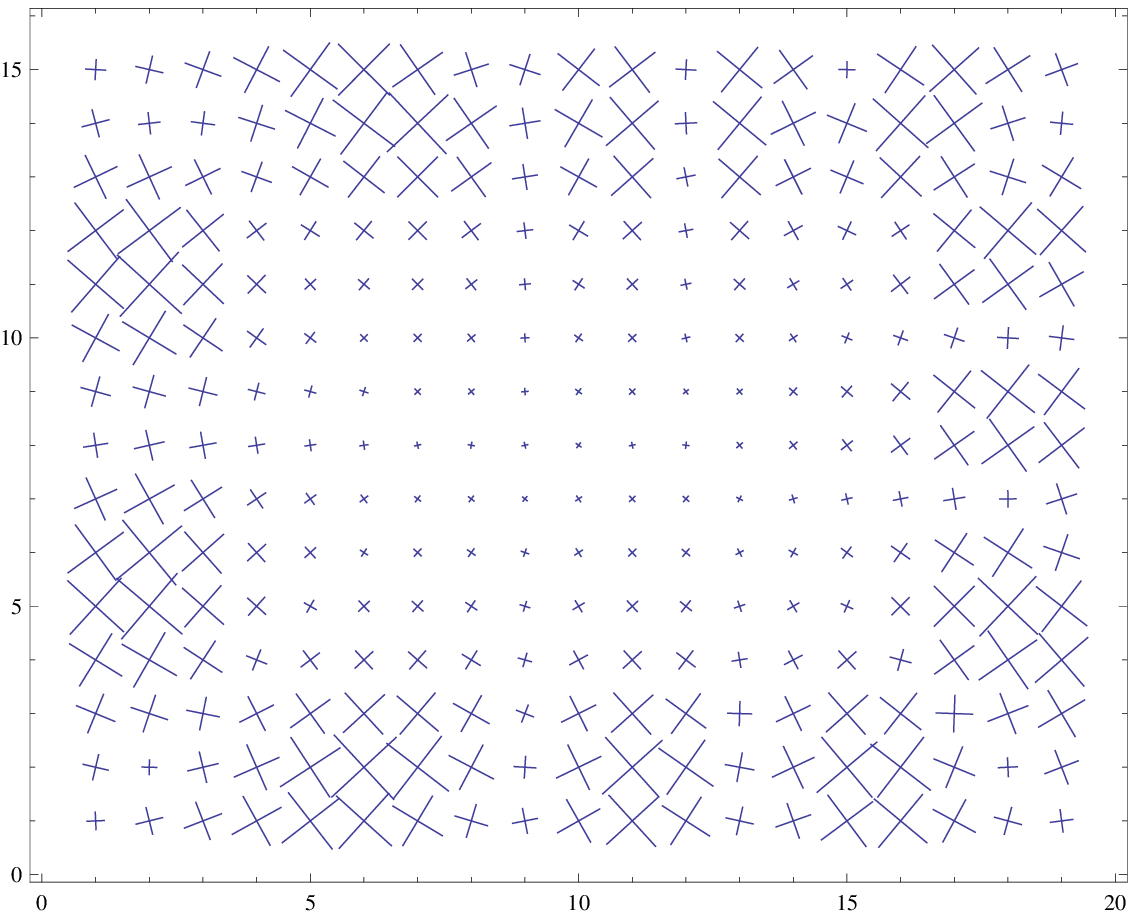}}
{\it Fig 3.} {Principal directions (lengths correspond to sizes of eigen values in logarithmic scale) for different degree:
(a) $d=0$, (b) $d=1$, (c) $d=2$, (d) $d=3$.
Lattice is $|\partial \Lambda \cup \Lambda| = 15\times19$
with 2 boundary layers.
Here the anisotropic Hamiltonian $H_{(1,10)}$ is considered with $\beta = 1$.}
\medskip
To compute the degree we use the discrete approximation based on finite-differences method:
$$\oint_\gamma M^{-1}(x) dM(x) \approx
\sum_{x_i\in \gamma} \bigg({M(x_i) \over \sqrt{|\det M(x_i) |}}\bigg)^{-1} 
\Big( {M(x_{i+1}) \over \sqrt{|\det M(x_{i+1}) |}} -
{M(x_i) \over \sqrt{|\det M(x_i) |}}\Big)\leqno(3.4)$$
and a similar formula for $- \oint_\gamma dM(x) M^{-1}(x).$
When angles are close to $\pi n / 2, \; n\in{\bf Z}$, eigenvalues of reduced vorticity matrices 
are close to zero due to properties of $\sigma^i_x(\theta_i)$ on the boundary.
Multiplying matrices with a big discrepancy in their eigenvalues
would lead to large computational errors;
to compensate for this effect we use in Eq.~(3.4) ``normalized'' matrices that are divided by square roots of their Jacobians.

The main factor of inaccuracy in degree calculations consists in the discrete approximation of the integral and the number of points 
on an integration contour.
As a rule, accurate results require a lot of points on the integration contour;
but for larger degrees the variation of the angle increases from point to point 
and so do the error due to discrete approximation.
This problem can be solved by enlarging the lattice size.
Fig.~4 presents vorticity patterns with different $\beta$'s 
for a lattice with $|\partial \Lambda \cup \Lambda| = 23\times33$ and two boundary layers.
\bigskip
\input epsf \centerline{(a)\epsfxsize=7truecm \epsfbox{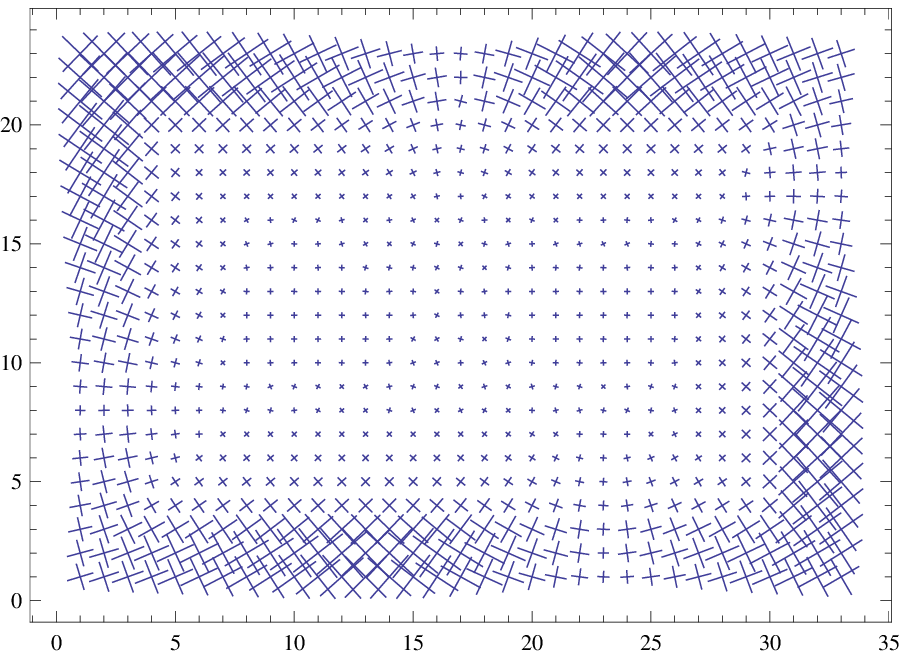}
\epsfxsize=7truecm \epsfbox{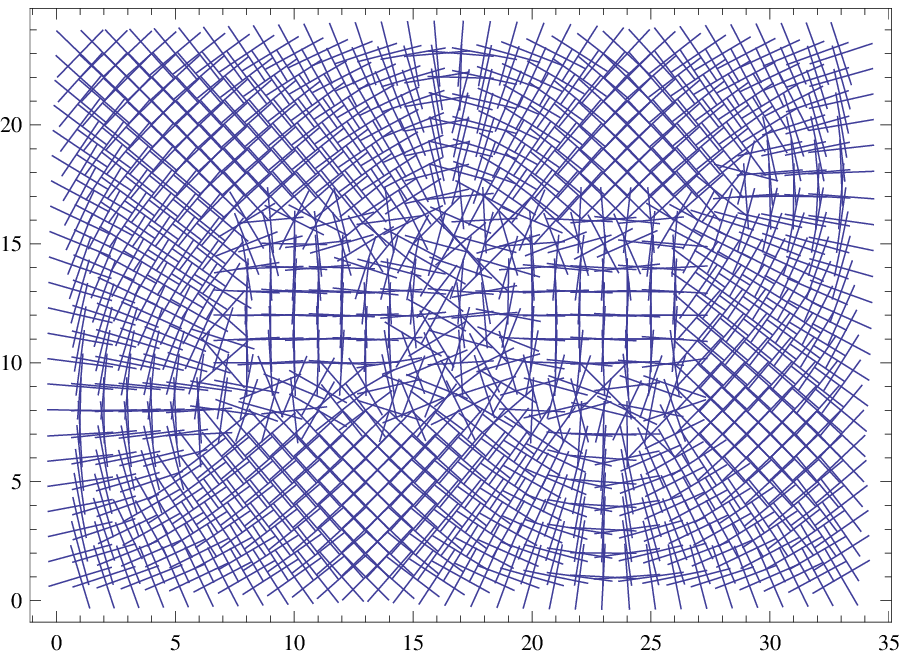}}
\input epsf \centerline{(b)\epsfxsize=7truecm \epsfbox{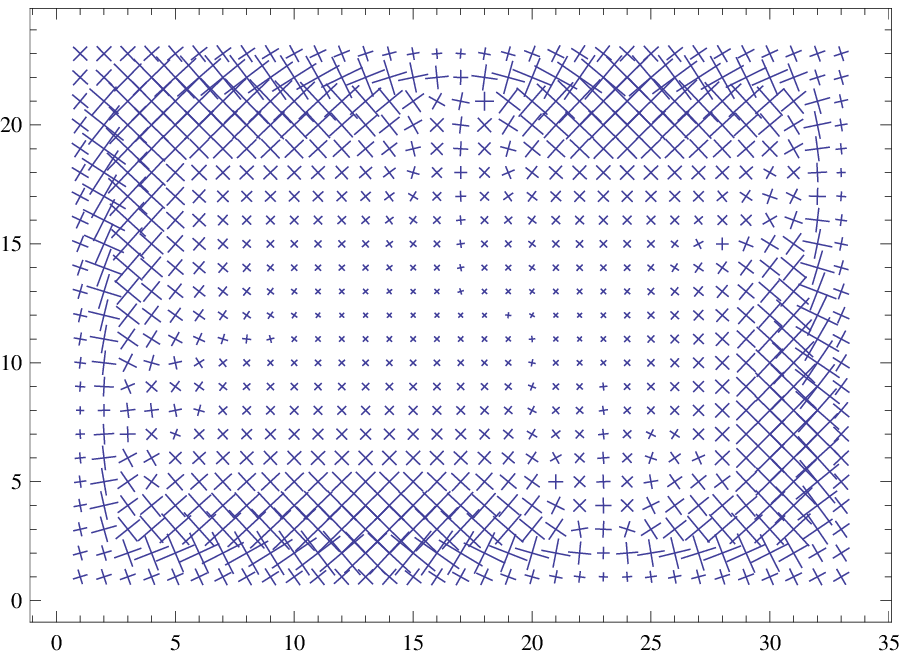}
\epsfxsize=7truecm \epsfbox{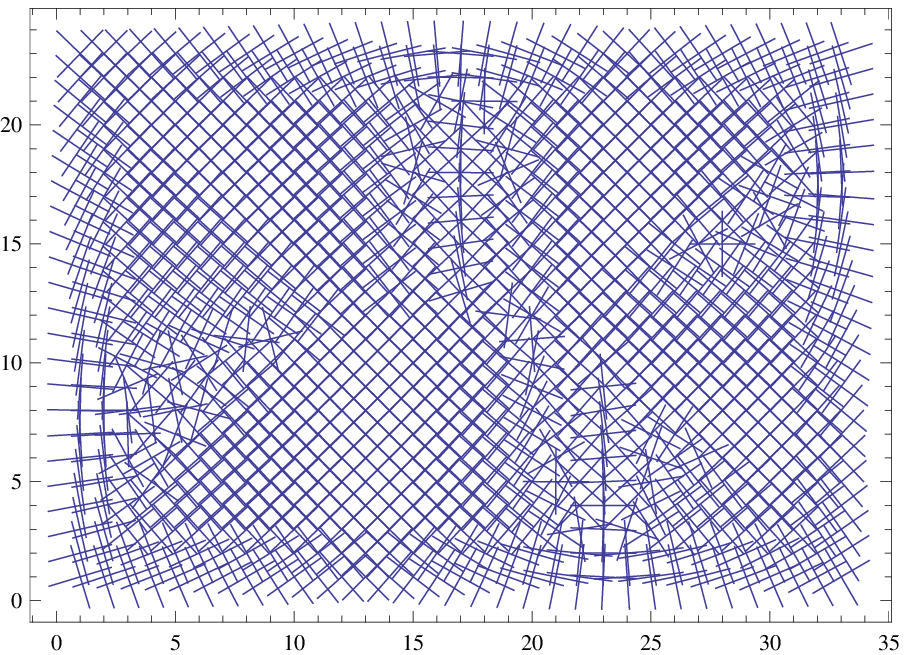}}
\input epsf \centerline{(c)\epsfxsize=7truecm \epsfbox{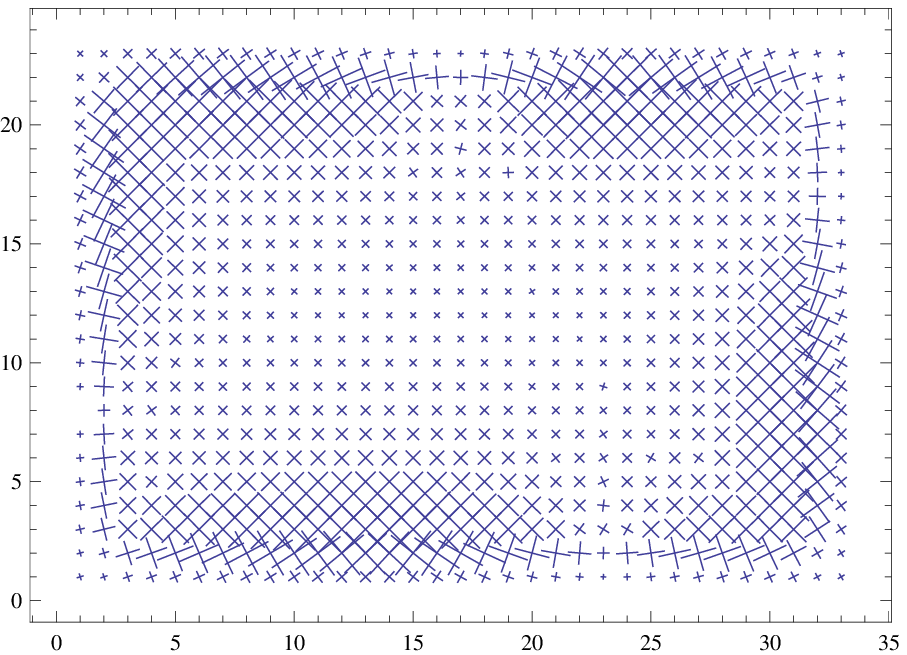}
\epsfxsize=7truecm \epsfbox{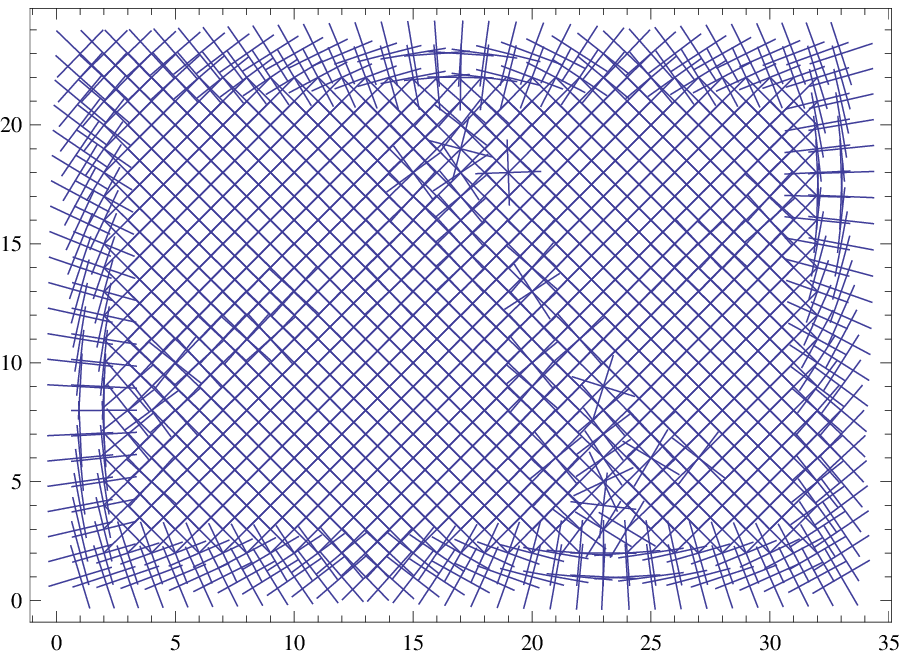}}
\medskip
{{\it Fig 4.} The principal directions for different beta: (a) $\beta=1$, 
(b) $\beta=8$, (c) $\beta=16$ for lattice $|\partial \Lambda\cup \Lambda| = 23\times33$ with 2 boundary layers.
Lengths of crosses correspond to eigenvalues on the logarithmic scale (left)
or are set equal (right).
Here the anisotropic Hamiltonian $H_{(1,10)}$ is considered.
The degree is $d=1$.
On the figure (a) to the right the eigenvalues in the center that are below the computational error are set to zero.}
\medskip 
Another factor that affects the computations is that eigenvalues are decaying while getting further from the boundary.
This we partially compensate by considering anisotropic model.
In principle we can consider anisotropic model with smaller anisotropy (e.g. $k=2$ or even close to 1).
As soon as $k>1$ eigenvalues inside the lattice are not zeros and the degree can be computed. 
But in practice for small anisotropy the rate of decay of eigenvalues is very high, so the computed degree is more accurate for large anisotropy. 
To make our vorticity patterns more demonstrative we consider the case of high anisotropy with $k=10$. In Table~1 
below we give the results for degree, computed for a large lattice along the cycle $\gamma\subset\Lambda$ 
consisting of the rectangle of the first or the second neighbors to the boundary for different values of the anisotropy parameter $k=2; 10$.
Inverse temperature is $\beta = 1$.
\medskip
{\it Table 1.} Table of calculated degree
for different values of anisotropy factor $k$
for the first and the second neighbors to the boundary.
Here the case of 2 boundary layers
$|\Lambda\cup\partial \Lambda| = 23\times 33$ is considered
with $\beta=1$.
$$\matrix{Given \,\, & k=2\,\, & k=10 \,\, & k=2\,\, & k=10\cr
degree &\, 1^{st} \, neighbours &\, 1^{st} \, neighbors & \, 2^{nd} \, neighbours 
& \, 2^{nd} \, neighbours   \cr
1&1.05&1.09&0.89&1.05\cr
2&1.98&2.03&1.70&1.78\cr
3&2.76&2.75&2.01&2.50 }
$$
\medskip
Increasing the boundary size for the same inner lattice does not almost change anything inside. This can be explained 
by the ``exponential'' decay of the information that propagates from boundary sites. 
So the influence of the third boundary layer on the inner points is too small 
compared with the effect of the closer sites on the first and the second boundary layers; in simulations it suffices to use only 2 boundary layers.

Let us consider the influence of the inverse temperature $\beta$.
As expected, the smaller $\beta$ the smaller are the eigenvalues of $\Omega^i_\beta(\delta)$ when $i\in\Lambda$,
because of disorder at high temperature;
but taking larger $\beta$ makes $\widehat\Omega^i_\beta(\delta)$, everywhere on $\Lambda$, very close to
$\pmatrix{0&-b_i\cr b_i&0\cr}$ which has
principal directions (1,1) and (1,-1), and the vorticity pattern is destroyed; 
this is not in contradiction with long range order at low temperature, but is probably due to volume effects.
(a) 
\medskip
Let us finally discuss the antiferromagnetic model. It is known that on ${\bf Z}^2$,
the unitary transformation $U$ consisting in flipping the spins at sites $i$
with $i$ odd (i.e. indices $i=(i_1,i_2)$ such that $|i|=|i_1|+|i_2|$ is odd)
intertwines the ferro with the antiferromagnetic models. More precisely, $-H=U^*HU$. The reason is that
${\bf Z}^2_e$ and ${\bf Z}^2_o$
(the even and odd lattices) are swapped into each other by symmetries on the lines
$x=n+1/2$ or $y=m+1/2$ (called the ``chessboard symmetry'').
There follows that $\tra \exp[\beta H]A=\tra \exp[-\beta H]UAU^*$, and if $A=\widetilde D^i$ (the canonical basis),
we can check $UAU^*=A$ so the matrices of vorticity (for the Hamiltonian with free boundary conditions) are the same.
This equivalence holds also in the case of the torus, but not on $\Lambda\subset{\bf Z}^2$ with an odd number of sites.
Of course, when $\partial\Lambda\neq\emptyset$, $H$ and $-H$ are not so simply related; nevertheless, we may observe (numerically) that the relation
$\Omega^i_\beta(\delta)=\Omega^i_{-\beta}(\delta)$ holds with a very good accuracy.
\medskip
\noindent {\it c) Summary}.
\smallskip
The number of vortices is equal to the topological degree and can be calculated by the integral along some contour. 
We use discrete approximation of this integral to compute the degree for a finite lattice $\Lambda\cup\partial\Lambda$. 
The main factors responsible for computational errors are the small number of points on the integration contour 
and the fast decay of eigenvalues inside the lattice.
We can use larger lattices and longer integration contour approaching the boundary to reduce the computational error 
related to the number of points, or
consider smaller degrees. We deal with anisotropic case to slow down the decay of eigenvalues near the center of $\Lambda$,
and take the inverse temperature of order 1.


\bigskip
\centerline{\bf Appendix}
\medskip
We provide a proof for the conjugacy relation (1.16) between $\delta$-symmetric basis.
Allow for the action of $G_0$, we can assume the matrix 
$\widehat b\in O(4)$ with columns $(b^1,b^2,b^3,b^4)$ already has determinant 1, so this is the matrix
of a rotation in ${\bf R}^4$. We recall from [So] the structure of matrices of rotation in ${\bf R}^4$.
Matrices of the  form 
$$A_G(\alpha,U)=\pmatrix{\cos\alpha&-\sin\alpha{}^tU\cr\sin\alpha U&\cos\alpha+\sin\alpha j(U)\cr}\leqno(a.1)$$
with ${}^tU=(x,y,z)$ a unit vector in ${\bf R}^3$, and 
$$j(U)=\pmatrix{0&-z&y\cr z&0&-x\cr -y&x&0\cr}$$ 
are called {\it left-quaternions}.  
In the same way, 
$$A_D(\alpha,U)=\pmatrix{\cos\alpha&\sin\alpha{}^tU\cr-\sin\alpha U&\cos\alpha+\sin\alpha j(U)\cr}\leqno(a.2)$$
are called {\it right-quaternions}. The point is that any $R\in O^+({\bf R}^4)$ can be written as the 
product of a left-quaternion and a right-quaternion. More precisely, there are real numbers
$\alpha, \beta$, and unit vectors $U=(\sin\theta\cos\phi,\sin\theta\sin\phi,\cos\theta)$, 
$V=(\sin\mu\cos\nu,\sin\mu\sin\nu,\cos\mu)$, such that $R=A_G(\alpha, U)A_D(\beta,V)$. Thus $R$ is a matrix 
depending on 6 variables $\alpha,\beta,\theta,\phi,\mu,\nu$. 

Recall from Definition 1.1 the conditions for $R$ to be a $\delta$-symmetric basis, $R\in[\delta]_s$; 
after taking some linear combinations between the original equations, this gives the following system:
$$\leqalignno{
\sin\,\beta \{\sin\,\mu [\cos\,\alpha (\cos\,\nu-&\sin\,\nu)
		+\sin\,\alpha\ \cos\,\theta (\cos\,\nu+\sin\,\nu)]\cr
	&-\sin\,\alpha\ \sin\,\theta\ \cos\,\mu (\cos\,\phi+\sin\,\phi)\}\cr
	&-\cos\,\beta\ \sin\,\alpha\ \sin\,\theta (\cos\,\phi-\sin\,\phi)=0&(a.3)\cr
&\sin\,\alpha\ \sin\,\beta\ \sin\,\theta\ \sin\,\mu\ \cos(\nu+\phi)=0&(a.4)\cr
&\sin\,\alpha\ \cos\,\beta\ \cos\,\theta + \cos\,\alpha\ \sin\,\beta\ \cos\,\mu=0&(a.5)\cr
&\sin\,\alpha\ \sin\,\theta [\sin\,\beta\ \cos\,\mu\ \cos\,\phi -\cos\,\beta\ \sin\,\phi]=0&(a.6)\cr
&\sin\,\beta\ \sin\,\mu [\cos\,\alpha\ \cos\,\nu +\sin\,\alpha\ \cos\,\theta\ \sin\,\nu]=0&(a.7)\cr
&\sin\,\alpha [2\,\cos\,\beta\ \cos\,\theta +\sin\,\beta\, \cos(\nu-\phi)\, \sin\,\theta\, \sin\,\mu] =0&(a.8)\cr
}$$
This system turns out to be overdetermined, which allows for discrete or 1-parameter families of solutions,
which can be found with the help of a symbolic computer program. 

Equation (a.4) yields the set of congruence mod $\pi$:
(a) $\alpha=0\, [\pi]$ or (b) $\beta=0\, [\pi]$ or (c) $\theta=0\, [\pi]$ or (d)$\mu=0 [\pi]$ or (e) $\nu+\phi={\pi\over2}\, [\pi]$.

In cases (a) or (b), we find $\alpha=0\, [\pi]$ and $\beta=0\, [\pi]$. In case (c), $\alpha=\beta={\pi\over2}\, [\pi]$
and $\mu=0\, [\pi]$. In case (d), $\mu=0 [\pi]$, $\alpha=0\, [\pi]$ and $\beta=0\, [\pi]$. 
So in cases (a) to (d), the solution to the system (a.3)-(a.8) consists in a discrete set.

Case (e) splits into two sub-cases:
(e') $\nu+\phi={\pi\over2}\, [\pi]$ with $\alpha\neq{\pi\over2}\, [\pi]$ and $\beta\neq{\pi\over2}\, [\pi]$, or
(e'') $\nu+\phi={\pi\over2}\, [\pi]$ with $\alpha\neq{\pi\over2}\, [\pi]$ or $\beta\neq{\pi\over2}\, [\pi]$.

In case (e'), we find $\bigl(\alpha={\pi\over4}\, [\pi] \ \hbox{and} \ \beta=-{\pi\over4}\, [\pi]\bigr)$, or
$\bigl(\alpha=-{\pi\over4}\, [\pi] \ \hbox{and} \ \beta={\pi\over4}\, [\pi]\bigr)$. 
So in cases (e'), the solution to the system (a.3)-(a.8) consists again in a discrete set.

In case (e'') we find the solutions we obtained before, in addition to a one-parameter set defined by 
$\alpha\equiv\beta\equiv\phi={\pi\over2}\, [\pi]$, $\nu=0\, [\pi]$, and $\mu=\theta=t\in{\bf R}$. 
Summing up:
\medskip
\noindent{\bf Lemma a.2}: 
{\it The solutions of $R\in[\delta]_s$ consist in a discrete set, comprising $R_0=\pm \Id$, and 
$$\eqalign{
&R_5=\pm\pmatrix{1 & 0 & 0 & 0 \cr 0 & -1 & 0 & 0 \cr 0 & 0 & -1 & 0 \cr 0 & 0 & 0 & 1\cr}, \ 
R_6=\pm\pmatrix{
	0 & 0 & 0 & 1 \cr 0 & 0 & -1 & 0 \cr 0 & -1 & 0 & 0 \cr 1 & 0 & 0 & 0\cr}\cr
&R_1= \pm{1\over2}\pmatrix{
	1 & 1 & 1 & 1 \cr -1 & 1 & -1 & 1 \cr -1 & -1 & 1 & 1 \cr 1 & -1 & -1 & 1\cr}, \
R_2= \pm{1\over2}\pmatrix{
	1 & -1 & 1 & -1 \cr 1 & 1 & 1 & 1 \cr -1 & 1 & 1 & -1 \cr -1 & -1 & 1 & 1\cr} \cr
&R_3=\pm{1\over2}\pmatrix{
	1 & 1 & -1 & -1 \cr -1 & 1 & 1 & -1 \cr 1 & 1 & 1 & 1 \cr -1 & 1 & -1 & 1\cr}, \
R_4= \pm{1\over2}\pmatrix{
	1 & -1 & -1 & 1 \cr 1 & 1 & -1 & -1 \cr 1 & -1 & 1 & -1 \cr 1 & 1 & 1 & 1\cr} \cr
}\leqno(a.9)$$
and 4 one-parameters families, with $0\leq s\leq1$~:}
$$\eqalign{
&R_{10}(s)=\pm\pmatrix{
	s & \sqrt{s(1-s)} & \sqrt{s(1-s)} & 1-s \cr \sqrt{s(1-s)} & -s & 1-s & -\sqrt{s(1-s)} \cr
	 \sqrt{s(1-s)} & 1-s & -s & -\sqrt{s(1-s)} \cr 1-s & -\sqrt{s(1-s)} & -\sqrt{s(1-s)} & s\cr}, \cr
&R_{11}(s)=\pm\pmatrix{
	-s & \sqrt{s(1-s)} & -\sqrt{s(1-s)} & 1-s \cr \sqrt{s(1-s)} & s & 1-s & \sqrt{s(1-s)} \cr
	 -\sqrt{s(1-s)} & 1-s & s & -\sqrt{s(1-s)} \cr 1-s & \sqrt{s(1-s)} & -\sqrt{s(1-s)} & -s\cr}\cr
&R_{12}(s)=\pm\pmatrix{
	-s & -\sqrt{s(1-s)} & \sqrt{s(1-s)} & 1-s \cr -\sqrt{s(1-s)} & s & 1-s & -\sqrt{s(1-s)} \cr
	 \sqrt{s(1-s)} & 1-s & s & \sqrt{s(1-s)} \cr 1-s & -\sqrt{s(1-s)} & \sqrt{s(1-s)} & -s\cr}\cr
&R_{13}(s)=\pm\pmatrix{
	s & -\sqrt{s(1-s)} & -\sqrt{s(1-s)} & 1-s \cr -\sqrt{s(1-s)} & -s & 1-s & \sqrt{s(1-s)} \cr
	 -\sqrt{s(1-s)} & 1-s & -s & \sqrt{s(1-s)} \cr 1-s & \sqrt{s(1-s)} & \sqrt{s(1-s)} & s\cr}\cr
}\leqno(a.10)$$
Now according to (1.5), we identify each column $b^k$ of these matrices with a $2\times2$ matrix $b_k$; 
it turns out that we can always relate the resulting $b$ with $\delta$ by the conjugacy relation (1.16) for some $P$. 
For instance, $R_{10}(s)$ gives the $\delta$-symmetric basis
$b=\pmatrix{b_1\kern 1pt &b_2\kern 1pt \cr
b_3\kern 1pt & b_4\kern 1pt }$ with 
$$\eqalign{
&b_1=\pmatrix{s\kern 1pt &\sqrt{s(1-s)} \kern 1pt \cr\sqrt{s(1-s)} \kern 1pt & 1-s}, \quad 
b_2=\pmatrix{\sqrt{s(1-s)}\kern 1pt &-s\kern 1pt \cr 1-s\kern 1pt &-\sqrt{s(1-s)}}\cr
&b_3=\pmatrix{\sqrt{s(1-s)}\kern 1pt &1-s\kern 1pt \cr -s\kern 1pt &-\sqrt{s(1-s)}}, \quad
b_4=\pmatrix{1-s\kern 1pt &-\sqrt{s(1-s)} \kern 1pt \cr-\sqrt{s(1-s)} \kern 1pt & s}\cr
}$$
and moreover, $b={}^tP_s\delta P_s$, with 
$$P_s=\pmatrix{\sqrt s\kern 1pt &\sqrt{1-s} \kern 1pt \cr\sqrt{1-s} \kern 1pt & -\sqrt s}$$
If we do no more allow for the action of $G_0$, we can look for a $\delta$-symmetric basis associated with an isometry $R\in O_-(4)$,
its product with an element of $G_0$ being again a rotation.
This readily gives (1.16) and Proposition 1.4 is proved.

\bigskip
\centerline{\bf References}
\medskip
\noindent [AsPi] W.Aschbacher, C.A.Pillet. Non-equilibrium steady states of the XY chain. J. Stat. Phys. 112, p.1153-1175, 2003.

\noindent [BePo] A.Belavin, A.Polyakov. Metastable tates of 2-d isotropic ferromagnets. JETP Lett., Vol 22, No.10, p.245-247, 1975.

\noindent [BiChSt] M.Biskup, L.Chayes, S.Starr. Quantum spin systems at finite temperature. Commun. Math. Phys.
Vol 269, 3, p.611-657, 2007. 

\noindent [BrFoLa] J.Bricmont, J.Fontaine, J.Landau. On the uniqueness of the equilibrium state for the plane rotator.
Commun. Math. Phys. 56, p.281-286 (1977)

\noindent [DaManTie] M.Damak, M.Mantoiu, R.Tiedra de Aldecoa. Toeplitz algebras and spetral results for the 1-D Heisenberg model.
J. Math. Phys. 47 (2006), no. 8, 082107

\noindent [De] O.Derzhko. Jordan-Wigner fermionisation for spin-1/2 systems in 2-D: a brief review. J. Phys. Studies (L'viv), 2000.


\noindent [DyLLiSi] F.Dyson E.Lieb, B.Simon.
Phase transitions in quantum spin systems with isotropic and  anisotropic interactions. J. Stat. Phys. 18(4),
p.335-383, 1978)
 
\noindent [El-BRo] H.El-Bouanani.
``Vortex et aimantation dans le mod\`ele de Kac.'' Ph.-D Thesis. Universit\'e de Toulon, Centre de Physique Th\'eorique (unpublished), 2008. 

\noindent [El-BRo] H.El-Bouanani, M.Rouleux. {\bf 1}. Vortices and magnetization in Kac's model. 
J. Stat. Physics. Vol.28, No.3, p.741-770, 2007. {\bf 2}. Thermodynamical equilibrium of vortices in
the continuous bidimensional Kac model (preprint arXiv 0707.226). 

\noindent [FrLi] J.Fr\"ohlich, E.Lieb. Phase transitions in anisotropic lattice spin systems. Commun. Math. Phys. 60, p.233-267 (1978)

\noindent [FrSp] J.Fr\"ohlich, T.Spencer. The Kosterlitz-Thouless phase transition in 2-D Abelian spin systems and the Coulomb gas.
Commun. Math. Phys. 81, p.527-602 (1981) 


\noindent [Ki] A.Kirillov. El\'ements de la Th\'eorie des Repr\'esentations, Editions MIR, Moscou, 1974. 


\noindent [Ma] P.Malliavin. G\'eom\'etrie Diff\'erentielle Intrins\`eque, Hermann, Paris, 1972.

\noindent [MeMiPf] A.Messager, S.Miracle, C.Pfister. Correlation inequalities and uniqueness of equilibrium state for the planar
rotator ferromagnetic model. Commun. Math. Phys. 58, p.19-29 (1978)

\noindent [Mi] J.Milnor. Topology from the differentiable viewpoint, Virginia Univ. Press, 1965.


\noindent [Ru] R.Ruamps. ``Mod\`eles de spins sur r\'eseau et vorticit\'e quantique''. Master's Thesis, 
Aix-Marseille Universit\'e, Centre de Physique Th\'eorique
(unpublished), 2008.

\noindent [ScHo] W.Schieve, L.Horwitz. Quantum Statistical Mechanics. Cambridge, 2009.

\noindent [Si] B.Simon. The Statistical Mechanics of Lattice Gas I. Princeton Univ. Press, 1993. 

\noindent [Sm] S.Smirnov. Discrete complex analysis and probability. 

http://www.unige.ch/~smirnov/slides/slides-hyderabad.pdf


\noindent [So] J.M.Souriau. Calcul lin\'eaire. PUF, Paris, 1965.

\bye